\documentclass[11pt,titlepage]{article}

\usepackage{setspace}
\usepackage[usenames,dvipsnames]{color}
\usepackage[normalem]{ulem}

\usepackage{amsmath,amssymb}

\usepackage{graphicx}
\usepackage{amssymb}
\usepackage[round,numbers,sort&compress]{natbib}
\def\be{\begin{equation}}
\def\ee{\end{equation}}
\def\bea{\begin{eqnarray}}
\def\eea{\end{eqnarray}}

\newcommand{\bdm}{\begin{displaymath}}
\newcommand{\edm}{\end{displaymath}}

\title{Force-induced dynamical properties of multiple cytoskeletal filaments are distinct from that of single filaments}

\author{
Dipjyoti Das\\
Department of Physics\\
 Indian Institute of Technology Bombay\\
 Mumbai 400076, India
\and
Dibyendu Das\\
Department of Physics\\
 Indian Institute of Technology Bombay\\
 Mumbai 400076, India
\and
Ranjith Padinhateeri\\
Department of Biosciences and Bioengineering\\
 Indian Institute of Technology Bombay\\
 Mumbai 400076, India\\
}

\date{}

\begin{document}

\maketitle
\begin{abstract}

{How cytoskeletal filaments collectively undergo growth and shrinkage is an intriguing question.
Collective properties of multiple bio-filaments (actin or microtubules)  undergoing hydrolysis, have not been studied extensively earlier, within simple theoretical frameworks. 
In this paper, we show that collective properties of multiple filaments under force are very distinct from the properties of a single filament under similar conditions -- these distinctions manifest as follows:  (i) the collapse time during collective catastrophe
for a multifilament system is much larger than that of a single filament with the same average length, (ii) force-dependence of the cap-size distribution of multiple filaments are quantitatively different from that of single filament, (iii) the diffusion constant associated with the system length fluctuations is distinct for multiple filaments (iv) switching dynamics of multiple filaments between capped and uncapped states and the fluctuations therein are also distinct. We build a unified picture by establishing interconnections among all these collective phenomena. Additionally, we show that the collapse times during catastrophes can be sharp indicators of collective stall forces exceeding the additive contributions of single filaments.
}
\end{abstract}


\section*{Introduction}

A large number of biological functions such as mitosis, acrosomal processes and cell motility are controlled by cytoskeletal filaments, whose classic examples are microtubules and actin filaments within cells \cite{howard-book}. Cytoskeletal filaments have different molecular structures --- the microtubule has a hollow cylindrical shape made of $13$ proto-filaments, while actin has helical shape made of two proto-filaments \cite{howard-book,Alberts-book}. In spite of their structural differences, these filaments have similar kinetic processes. They  polymerize by adding ATP/GTP-bound subunits.  Inside a filament,  ATP/GTP is irreversibly hydrolysed into ADP/GDP. The presence of this chemical switching (ATP/GTP hydrolysis) makes the growth dynamics {\it non-equilibrium} in nature, and produces two distinct subunit-states, namely ATP/GTP-bound and ADP/GDP-bound. These two subunit-states have very distinct depolymerization rates, and this heterogeneity  produces interesting dynamics \cite{Pollard1986,Desai_Mitchison_MT:97}.

In the literature, the dynamics of a single cytoskeletal filament has been studied extensively \cite{howard-book,hill:81, kolomeisky:06, Leibler-cap:96,vavylonis:05,Ranjith2009,Ranjith2010,Brun-2009,Ranjith2012,anderson:2013}.  Single microtubules are known to exhibit a phenomenon called ``dynamic instability'' where the filament grows with a certain velocity, and then collapses catastrophically generating a huge fluctuation in the filament lengths~\cite{mitchison:1984, Desai_Mitchison_MT:97}. It has been reported that single actin filaments and ParM filaments (homologue of actin in prokaryotes) also exhibit large length fluctuations, somewhat similar to microtubules~\cite{fujiwara:02, Garner-etal_Sci:04}. Given that these filaments bear load under various circumstances, scientists have also investigated how these filaments and their length fluctuations behave under force~\cite{jason-dogterom:03}.

Extensive theoretical investigation, combined with experiments, have given us a good primary understanding of how these filaments behave at the single filament level. Early  phenomenological models tried to capture the filament dynamics by a two-state model \cite{Leibler:93} with stochastic transitions between growing and shrinking length-states.  Later  models incorporated detailed chemical processes such as binding and unbinding of monomers, and hydrolysis, using experimentally  measured rates \cite{vavylonis:05,Ranjith2009,Ranjith2010,Ranjith2012}. All these studies revealed that the chemical switching (hydrolysis) is crucial to explain the experimentally observed  feature  of  ``dynamic instability'' \cite{Desai_Mitchison_MT:97,howard:2009} and similar large length fluctuations~\cite{vavylonis:05}.  
The reason behind this fluctuation phenomenon was found to be the formation of a ATP/GTP-cap at the filament-tip 
and {the stochastic} disappearance of it due to hydrolysis.  

Although single-filament studies are  helpful to understand the basic aspects of the dynamics, it is biologically more relevant to investigate a collective system of $N (>1)$ filaments. {Even though scientists are starting to explore dynamics of multiple filaments under force experimentally~\cite{footer-dogterom:07,Laan-pnas:08}, the theoretical understanding of multi-filament dynamics and their fluctuations is minimal.} 
Most of the existing models for multi-filaments neglect ATP/GTP hydrolysis and do not have any kind of chemical switching in their model~\cite{kolomeisky:05, van-doorn:00,lacoste11, Kierfeld:2011,ramachandran:2013}. 
Ignoring hydrolysis, for  simple models of filaments with polymerization and depolymerization dynamics, exact analytical results for $N=2$ \cite{lacoste11,kolomeisky:05,Kierfeld:2011}, and numerical results  for $N\geq2$ \cite{van-doorn:00,lacoste11, Kierfeld:2011,ramachandran:2013} have been obtained.
Given that single-filament studies have already established the  experimental importance of chemical switching~\cite{hill:85,Leibler-cap:96,kolomeisky:06, vavylonis:05}, it is crucial to have a multi-filament study where one takes into account the ATP/GTP hydrolysis in detail and investigate the dynamics. Also note that the irreversible process of hydrolysis makes the dynamics depart from equilibrium, and hence it needs careful consideration.

 In the context of force generation, in a recent study, we have theoretically shown that ATP/GTP hydrolysis results in a new collective phenomenon~\cite{das:arxiv}.  For a bundle of $N$ parallel filaments pushing against a wall, the collective stall force  is {\it greater} than $N$ times the stall force of a single filament~\cite{das:arxiv}. 
 Earlier theories \cite{van-doorn:00,lacoste11} missed this effect as they neglected hydrolysis and studied equilibrium processes, which led to a notion that stall forces are additive for multiple filaments.

%
Apart from force generation, various fluctuations of the system-length during unbounded growth or 
``catastrophes'' have been of great interest \cite{Laan-pnas:08,vavylonis:05,Ranjith2010,Ranjith2012}. Single-filament studies have described the  length fluctuations by a  measurable quantity, namely the diffusion constant \cite{vavylonis:05,Ranjith2010,kolomeisky:05,kolomeisky:06}. {Recent theoretical studies of  single actin filaments have shown that this diffusion constant has non-monotonic behavior as a function of monomer concentration \cite{vavylonis:05,kolomeisky:06}} -- it has a peak near the critical concentration. It should be noted that such a peak would be absent without hydrolysis, which makes the filament switch between ATP/GTP ``capped'' and ``uncapped'' states \cite{vavylonis:05}.  
Another aspect of length fluctuation is the catastrophe and rescue  
where the filament repeatedly grows and shrinks maintaining a constant average length~\cite{howard-book}. Such stochastic length collapses recently have been observed for  multiple microtubules in an experiment \cite{Laan-pnas:08}, and have been referred to as ``collective catastrophes''. 
 
A unified theoretical understanding of the above fluctuation properties ({diffusion constant, catastrophes and cap dynamics}) have not been provided in any earlier literature for multiple filaments under force, and undergoing hydrolysis. {Zelinski and Kierfeld have theoretically studied the collective catastrophe using a phenomenological two state model~\cite{Kierfeld:2013}. However, none of the existing multifilament models take into account microscopic processes like polymerisation, ATP/GTP hydrolysis and depolymerisation of ATP/GTP- and ADP/GDP-bound subunits explicitly. Given that explicit dynamics at the subunit level is crucial in understanding the coupling between cap dynamics and length fluctuations},  
it is desirable to have a microscopic model that includes these features in detail.
%
%
%

Motivated by the above research background, in this paper we investigate the dynamics of multiple cytoskeletal filaments {taking into account the kinetic events of polymerisation, depolymerisation, and ATP/GTP hydrolysis of subunits explicitly. The focus of the paper is to examine the collective properties that may emerge from the multifilament nature of the system, in the presence of force and non-equilibrium ATP/GTP hydrolysis. We show that collective behaviour of multi-filaments under force is qualitatively and quantitatively different from that of a single filament, and the ATP/GTP cap dynamics is crucial in understanding these phenomena. Examining the collapse during catastrophe, we show that the collapse time of a multifilament system is considerably higher than that of a single filament system; this indicates that the collective collapse of microtubules has a gradual nature as opposed to the sharp collapse of single microtubule}. We find that this slow collapse of the multi-filament system is related to the enhanced stability of the ATP/GTP caps. We establish this by studying the cap-size statistics, and the switching dynamics of the system between capped and cap-less states. {We find that the multifilament system has a non-zero cap, at any large force, while for a single filament cap vanishes at large forces}. Finally, we show that these underlying features manifest in the macroscopic fluctuations of the system size {and can be quantified 
as the experimentally measurable diffusion coefficient. Through this paper, we provide a unified picture by establishing connections between a number of collective properties of the multifilament system and the underlying kinetics of the AGP/GTP cap at the subunit level.}
\begin{figure}[ht]
\centering
\includegraphics [angle=0,scale=0.4]{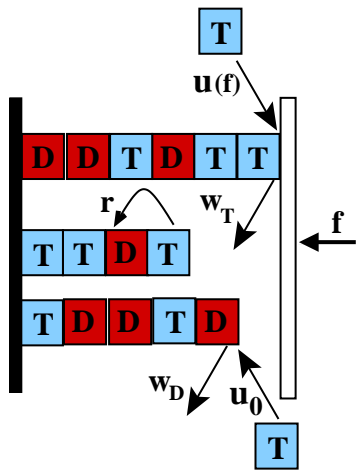}
\caption{ Schematic diagram of three-filament system with random hydrolysis, where the switching ATP/GTP $\rightarrow$  ADP/GDP  occurs  randomly at any ATP/GTP subunit. ATP/GTP and ADP/GDP subunits are shown as letters `T' (blue) and `D' (red) respectively. The left wall is fixed, while the right wall is movable with an externally applied force $f$ pushing against it. Various possible events (as described in the text) are shown with arrows and corresponding rates.
}
\label{fig1}
\end{figure}

\section*{Model}

We study a model of multiple cytoskeletal filaments (see Fig. \ref{fig1}), where $N$ parallel and rigid filaments (actins or microtubules), each composed of subunits of length $d$, are growing against a wall under a constant opposing force $f$. This model is a generalisation of the one-filament random hydrolysis model \cite{Ranjith2012} to a multi-filament case. In this model, 
each filament grows by polymerisation of free ATP/GTP-bound subunits in a force-dependent manner. Filament tips away from the wall polymerise with a
rate $u_0=k_0 c$. Here, $k_0$ is the intrinsic polymerization rate-constant and $c$ is the free ATP/GTP subunit concentration. The polymerization rate for the leading filament, which is in contact with the wall, is reduced due to the  applied force $f$ --- according to the Kramer's theory, the rate becomes  $u(f)=u_0 {\rm e}^{-f d/K_B T}$ \cite{lacoste11,van-doorn:00}. Inside each filament, any ATP/GTP-bound subunit may get hydrolysed  to a ADP/GDP-bound subunit randomly at any location with a rate $r$. This random mechanism of hydrolysis is thought to be closer to the biological reality \cite{Jegou2011}. Note that the chemical switching  (ATP/GTP $\rightarrow$ ADP/GDP) 
is   {\it non-equilibrium} in nature, 
as it is irreversible. Finally, the ATP/GTP-bound and ADP/GDP-bound subunit may dissociate from the tip of a filament with distinct force-independent depolymerization rates $w_T$ and $w_D$ respectively. The continuous ATP/GTP stretch at the tip of a filament is called a ``cap'' --- for example, in Fig. \ref{fig1}, the top filament has a cap whose size is two subunits. Note that the immovable left wall (see Fig. \ref{fig1}) acts as a reflecting boundary --- this is equivalent to a filament growing from a fixed seed on the wall, where the filament
can polymerise back once its length reduces to zero.  
  In this coarse-grained  model, the effective subunit lengths are taken to be $d=5.4 \rm{nm}/2=2.7 \rm{nm}$ for actin filaments, and $d=8 \rm{nm}/13=0.6 \rm{nm}$ for microtubule, which accounts for the actual multi-protofilament nature of the biofilaments \cite{Ranjith2009,Ranjith2010,Ranjith2012,Kierfeld:2013}. We do kinetic Monte-Carlo simulations  \cite{gillespie:77} of the above model using known rates for cytoskeletal filaments (see table \ref{table1}) to calculate various dynamical quantities, and the results are given below.
  
\begin{table}
\caption{Rates for Actin~\cite{Pollard1986, howard-book} and Microtubules (MT) ~\cite{Desai_Mitchison_MT:97, howard-book, mitchison:1984}}
\label{table1}
\centering
\begin{tabular}{|c|c|c|c|c|c|} \hline
~ & $k_0$ ($\mu M^{-1}s^{-1}$) & $w_T ~(s^{-1}) $ & $w_D ~(s^{-1}) $ & $r$ ($s^{-1}$) \\ \hline
Actin & $11.6$ & $1.4$ & $7.2$  & $0.003$ \\ \hline
MT & $3.2$ & $24$ & $290$  & $0.2$ \\ \hline
\end{tabular}
\end{table}


\section*{Results}

\section*{Collapse times reveal novel collective behaviour during catastrophe under force}
%
%

In this section, we study the collective collapse of $N$ filaments during catastrophes.  We simulate an $N$ filament system growing against a wall under external force $f$, as discussed above. When the external force is larger than the ``stall force'' (maximum force) of the N-filament system ($f_s^{(N)}$), the filaments will not grow on an average -- the system will be in a bounded phase of growth (see Appedix A).

\begin{figure}[ht]
\centering
\includegraphics [angle=-90,scale=0.4]{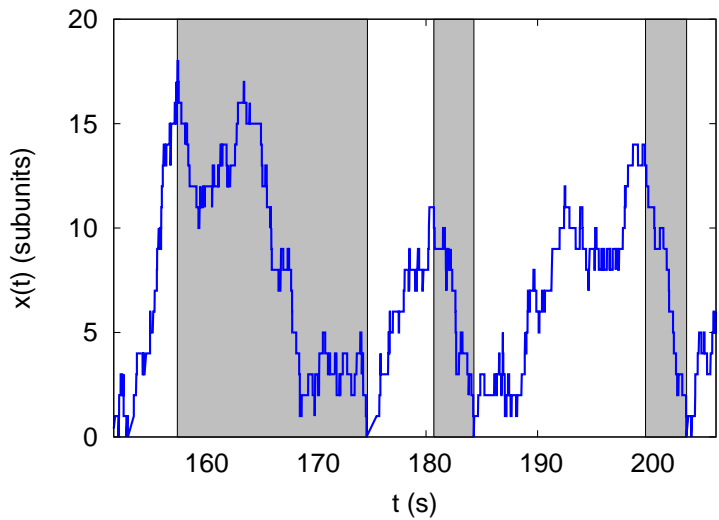}
\caption{A time trace of the wall position $x(t)$ for two microtubules $(N=2)$ in the bounded phase, showing ``collective catastrophe'', at a concentration $c=100 \mu$M$\gg c_{crit}$, and at a force $f=36.8$ pN$> f_s^{(2)}$ ($c_{crit}=8.67 \mu$M, and $f_s^{(2)}=35.0$pN in this case). Other parameters are taken from Table 1. The  regions shaded grey correspond to the catastrophes, and provide the collapse time intervals whose average is  $T_{\rm{coll}}$.}
\label{fig2}
\end{figure}

First of all, our model shows collective catastrophes of multiple filaments in the bounded phase,  similar to a recent experiment \cite{Laan-pnas:08}. A typical time trace of the wall position (or equivalently system-length) is given in Fig. \ref{fig2}, where a system of two microtubules repeatedly grows from a length of zero to a maximum value and then shrinks back to zero. This stochastic collapses of the system-length from a local maximum to zero, would be referred to as ``catastrophes''.  To quantify and
systematically investigate the catastrophe events, we define a measurable quantity called collapse time below: following Fig. \ref{fig2}, we define a  ``peak" as the furthest wall position between two successive zero values of the system-length ($x$). Then we define the collapse time ($T_{\rm coll}$) as the time it takes, on an average, to collapse from a peak to the next zero of the system-length (see the  regions shaded grey in Fig. \ref{fig2}). Below stall force, where the system would be in a unbounded growing phase (see Appendix A), the  $T_{\rm coll}$, according to our definition, would be infinite as the trajectories of the system-length would no longer collapse to zero (on an average).
In other words, $T_{\rm coll}$ is expected to diverge for $f \le f_s^{(N)}$. On the other hand, $T_{\rm coll}$ should be  finite in the bounded phase (see Fig.~\ref{fig2}) as there are frequent catastrophes. Thus, the finiteness of $T_{\rm coll}$ values is a quantitative indicator of the existence of catastrophes.  
\begin{figure}[ht]
\centering
\includegraphics [scale=0.45]{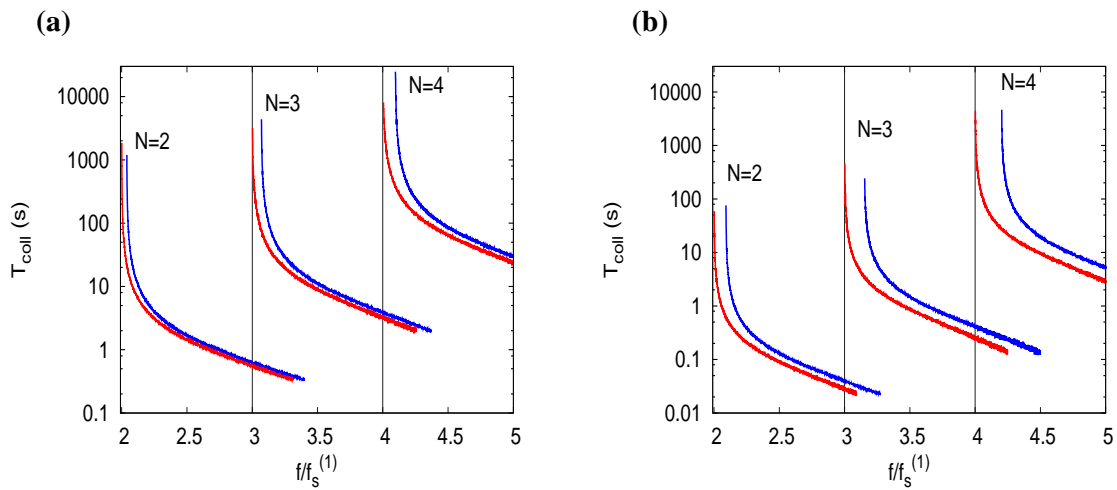}
\caption{Average collapse times $T_{\rm coll}$ as a function of scaled force $f/f_s^{(1)}$ with increasing number of filaments ($N$), for (a) actins and (b) microtubules. Blue and red curves are  with hydrolysis ($r>0$) and without hydrolysis ($r = 0$) respectively. The curves are plotted by scaling the force-axis with corresponding single-filament stall forces. For $r > 0$, the  numerically obtained values  of single-filament stall forces are $f_s^{(1)}=3.13$ pN for actin, and  $f_s^{(1)}=16.75$ pN for microtubule. While, for $r=0$, the corresponding single-filament stall forces are obtained from the formula $f_s^{(1)}= (k_B T/d)~ \rm{ln}(k_0 c/w_T)$ (see \cite{van-doorn:00}) --- these are $f_s^{(1)}=3.21$ pN for actin, and  $f_s^{(1)}=17.70$ pN for microtubule. Parameters are taken from table \ref{table1}. The ATP/GTP concentrations are $c=1\mu M$ for actin, and  $c=100\mu M$ for microtubules.}
\label{fig3}
\end{figure}

In Fig. \ref{fig3}, we plot  $T_{\rm coll}$ as a function of scaled force $f/f_s^{(1)}$, for multiple actin filaments (Fig.~\ref{fig3}a, blue curves) and microtubules (Fig.~\ref{fig3}b, blue curves). As expected, at large forces, the values of $T_{\rm coll}$ are finite, corresponding to the bounded phase. However, they diverge at specific  force values which are nothing but the collective \emph{stall forces} of N filaments $f_s^{(N)}$. Interestingly, we see that $f_s^{(N)} > N f_s^{(1)}$. This collective phenomenon of excess stall force generation (as opposed to $f_s^{(N)} = N f_s^{(1)}$) was recently  discovered by us  \cite{das:arxiv}; we had obtained $f_s^{(N)}$ by computing the force at which $\langle V \rangle \rightarrow 0$ (see \cite{das:arxiv}). 
Note that here we are estimating $f_s^{(N)}$ from the $f>f_s^{(N)}$ regime (bounded growth phase), while in \cite{das:arxiv}, the approach was from the $f<f_s^{(N)}$ regime (unbounded growth phase) -- see Appendix B for a comparison. It is important to stress that if hydrolysis is ignored, i.e. for the hydrolysis rate $r=0$, one obtains the red curves in Fig. \ref{fig3} --- they show $f_s^{(N)} = N f_s^{(1)}$, a widely believed result, but nevertheless actually untrue in reality.

\begin{figure}[ht]
\centering
\includegraphics [scale=0.45]{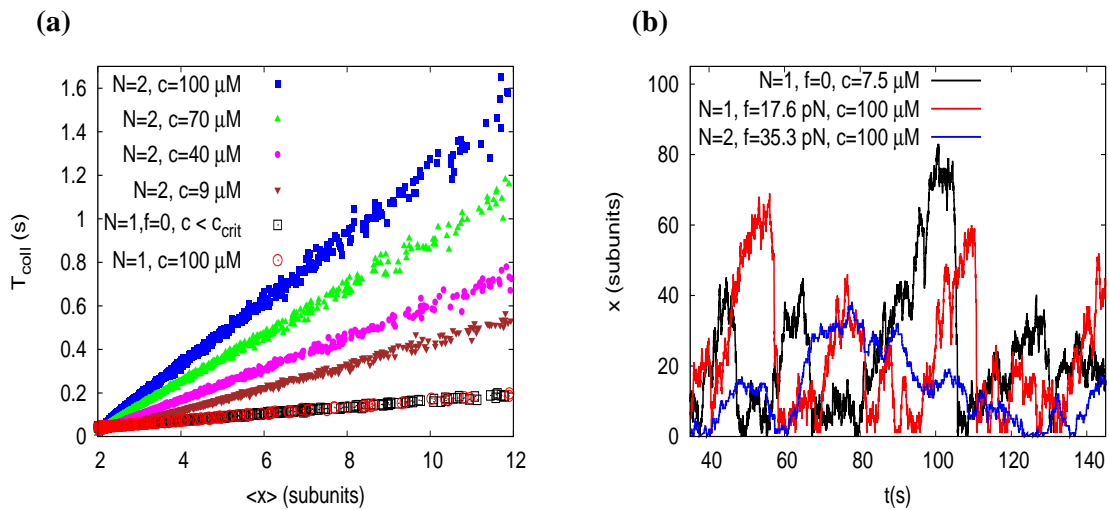}
\caption{ (a) Plot of $T_{\rm coll}$ versus $\langle x\rangle$ obtained in the bounded phase, for  microtubules. Black squares ($\Box$) are for $N=1$, $f=0$, $c<c_{crit}$, and red circles (\textcolor{red}{$\bigcirc$}) are for  $N=1$, $f>f_s^{(1)}$, $c=100 \mu$M$\gg c_{crit}$ (for microtubule parameters, $c_{crit}=8.67 \mu$M). While, for $N=2$, $f>f_s^{(2)}$,  values of $T_{\rm coll}$  are obtained at four different concentrations (all greater than $c_{crit}$)  $c=9 \mu$M (\textcolor{Brown}{$\blacktriangledown$}),  $40 \mu$M (\textcolor{Magenta}{$\bullet$}),  $70 \mu$M (\textcolor{Green}{$\blacktriangle$}), and $100 \mu$M (\textcolor{Blue}{$\blacksquare$}).  (b) A comparison of three time-traces of the wall-position $x(t)$ in the bounded phase, for parameters: (i) $N=1$, $f=0$, $c=7.5 \mu$M$<c_{crit}$ (black curve); (ii) $N=1$, $f=17.6$ pN$>f_s^{(1)}$, $c=100 \mu$M (red curve); and (iii) $N=2$, $f=35.3$ pN$>f_s^{(2)}$, $c=100 \mu$M (blue curve). Note that $\langle x\rangle$ is nearly same for all trajectories. }
\label{collapse}
\end{figure}

As $T_{\rm coll}$ is a nice quantitative measure of catastrophes, 
we would like to use it to address two questions: (a) is nature of the catastrophe of multiple filaments (collective catastrophe)  different from that of a single filament? (b) is there any difference between zero-force catastrophe and force-dependent collective catastrophe? Before proceeding to answer these two questions, we note that two external factors can control the catastrophe -- force and concentration of subunits (see Appendix A). In the absence of any force, all filaments are independent of each other, and therefore the average behaviour of N filaments is exactly the same as that of a single filament.
However, in the presence of force, the filaments interact via the movable wall. Due to the applied force, the growth rate of a filament, which is otherwise $k_0 c$, reduces instantaneously to $u(f)=k_0 c \rm{e}^{-f d/K_B T}$,  the moment it touches the wall. By this mechanism the trailing filaments get affected by
the spatial location of the leading filaments. This implicit interaction among filaments for $f>0$, can potentially lead to  
new collective phenomena for multi-filament systems, as we would show soon. 

 Noting these points, we proceed to compare the catastrophes for  the following three cases: (i) $N=1$, $f=0$, $c<c_{crit}$, (ii) $N=1$, $f>f_s^{(1)}$, $c>c_{crit}$ and (iii) $N=2$, $f>f_s^{(2)}$, $c>c_{crit}$.
Since the parameter regimes of the three different cases are very distinct, we present a scatter plot (see Fig. \ref{collapse}a) between the collapse time ($T_{\rm coll}$) and the average length of the leading filament (or the mean wall position).
Firstly we see that for a single filament ($N=1$), the $T_{\rm coll}$ data for the case (i) (by varying $c$), and for the case (ii) (by varying $f$), completely collapse on to each other (see bottom curves with symbols  $\Box$ and \textcolor{red}{$\bigcirc$} in  Fig. \ref{collapse}a). This means that the average collapse times of a single filament with or without force are similar. But, the situation is strikingly   different for $N>1$ in presence of a force, as we see below.

For $N=2$ microtubules (case (iii)), we calculated the values of $T_{\rm{coll}}$  at four different concentration values greater than $c_{crit}$ (blue, green, magenta and brown symbols in Fig. \ref{collapse}a) by varying forces $f>f_s^{(2)}$. We clearly see that  the values of $T_{\rm{coll}}$ are much higher compared to those of $N=1$, for the same given average length.  This implies that, during catastrophes of $N>1$ filaments under force, the system-length collapses more {\it slowly}, than a single filament. This behavior can be further seen in  Fig. \ref{collapse}b, where we show comparative time histories of the wall position for all three cases (i), (ii) and (iii). We see sharp length collapses for $N=1$ (for both cases (i) and (ii)), and comparatively much gradual catastrophes for $N=2$ (case (iii)).   
Above observations clearly indicate that, the system of multiple filaments under force seem to be more ``stable'' in comparison to a single filament during their catastrophes in the bounded phase. By ``stability'' we mean that multiple filaments resist the opposing force more effectively and thus collapse more slowly compared to $N=1$. 

\begin{figure}[ht]
\centering
\includegraphics [scale=0.55]{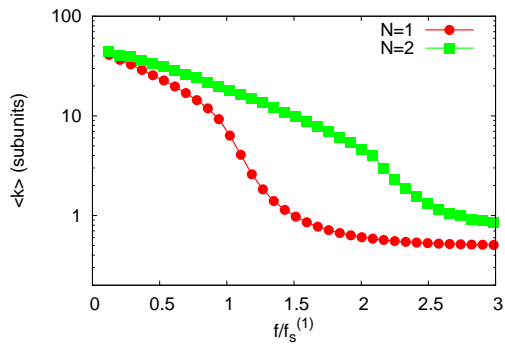}
\caption{ Average cap size $\langle k \rangle$  as a function of scaled force $f/f_s^{(1)}$ for microtubules, and for filament numbers $N=1$ (red), and  $N=2$ (green). The system is in the bounded phase for forces greater than the stall forces. The GTP concentration is  $c=100 \mu M$, and other parameters are specified in Table \ref{table1}. The Y-axis is in log scale.}
\label{cap_bound}
\end{figure}

Sudden collapse, during catastrophe, is typically associated with the disappearance  of ATP/GTP cap and exposure of ADP/GDP bulk, while the stability is associated with the presence of the ATP/GTP cap. This raises an obvious question:
Do  slow collapses during collective catastrophe, exhibited by the multi-filament system, have something to do with ATP/GTP cap stability?   To get a preliminary understanding,  we calculated the average cap sizes $\langle k \rangle$  as a function of force, for $N=1$ and $N=2$ in the bounded phase --- this is shown in Fig. \ref{cap_bound}.  This figure clearly shows that average cap sizes of a two-filament system is greater than that of a one-filament system. This points to a new  cap structure for collective ($N>1$) dynamics. In the next section, we examine these collective effects on  cap size statistics and cap dynamics in detail.

%
%
%


\subsection*{Multiple filaments under force show distinct cap-size statistics}

In this section, we study the statistics of ATP/GTP cap-sizes with the aim of understanding how it renders stability to a multi-filament system and slows down the catastrophe. Since our goal is to understand the steady-state properties of the caps during catastrophe, we start with  very long filaments. By studying the shrinkage of such
filaments we can examine the collective behaviour of their caps, without any 
boundary effect that may arise from the rigid wall at zero length.

\begin{figure}[]
\centering
\includegraphics [scale=0.45]{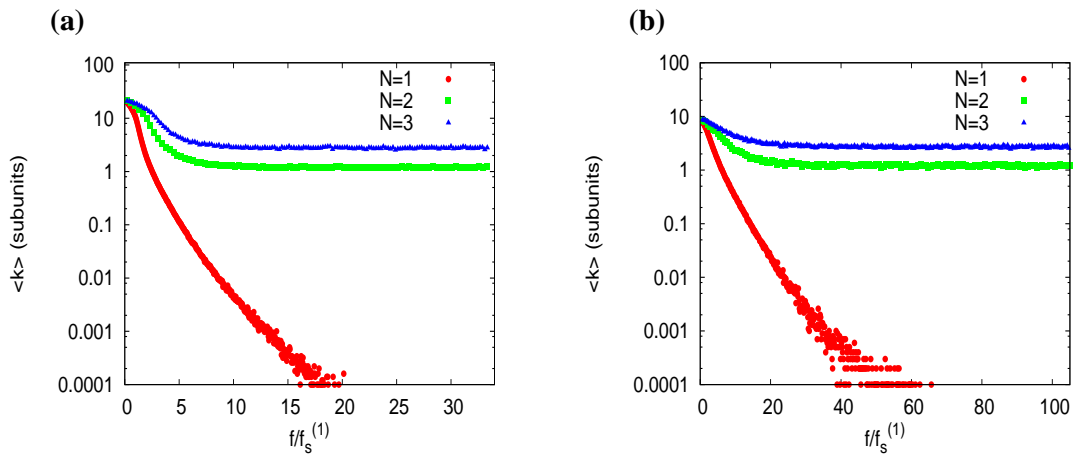}
\caption{Average cap size $\langle k \rangle$  as a function of $f/f_s^{(1)}$ for (a) actins and (b) microtubules, and for filament numbers $N=1$ (red), $N=2$ (green) and $N=3$ (blue). The concentrations are $c=0.2\mu M$ for actins, and  $c=10\mu M$ for microtubules. Y-axes are in log scale.}
\label{ave-cap}
\end{figure}

In Fig. \ref{ave-cap}, we plot  $\langle k \rangle$ against the scaled force $f/f_s^{(1)}$,  for actin filaments   (Fig. \ref{ave-cap}a) and microtubules (Fig. \ref{ave-cap}b).  Note that this figure is  the counterpart of Fig. \ref{cap_bound}, that  was studied for short filaments with possible boundary effects (see previous section). In Fig. \ref{ave-cap}, when $f \gg f_s^{(1)}$, we see that mean cap-length $\langle k \rangle$, for single filament, rapidly decays to zero (see red curves in Fig. \ref{ave-cap}). But for $N>1$, $\langle k \rangle$ does not vanish  at all --- rather, it first reduces and then {\it saturates} (or stabilizes) to a finite value of $\gtrsim 1$ subunits, at forces  $f \gg f_s^{(N)}$ (see green curves for $N=2$, and blue curves for $N=3$ in  Fig. \ref{ave-cap}).  These results reaffirm our observation in the last section that the multifilament system does show a distinct cap structure -- 
while average cap length of a single filament is vanishingly small, the multifilament system always has a non-vanishing larger cap. Does this also reflect in the full cap size distribution?
%

\begin{figure}
\centering
\includegraphics [scale=0.45]{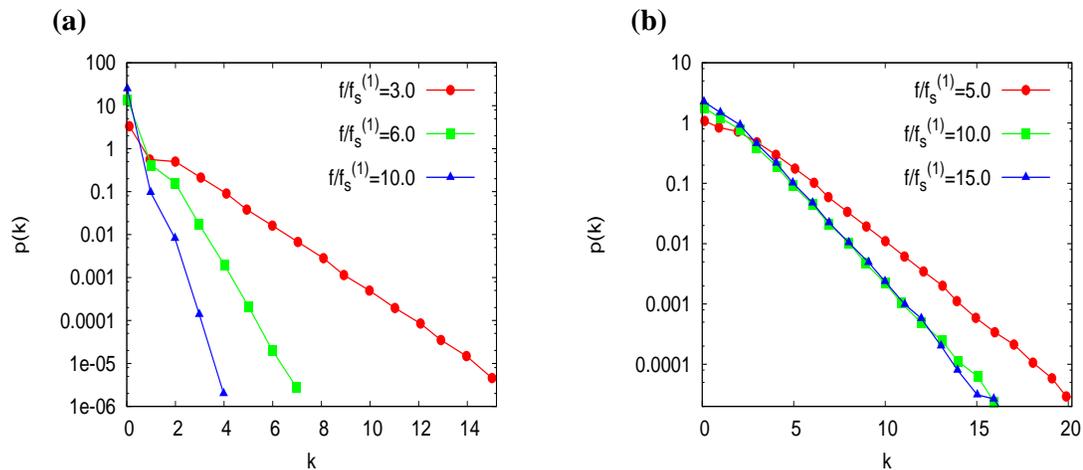}
\caption{ Distributions of cap sizes, $p(k)$ at different forces for (a) single actin ($N=1$), and (b) two actin filaments ($N=2$), for a  concentration $c=0.2\mu M$. Y-axes are in log scale.}
\label{cap-dist}
\end{figure}

In Fig. \ref{cap-dist}a, we plot the cap-size distributions $p(k)$ for a single actin filament at three different force values. We clearly see that the cap-size distributions for $N=1$ have decreasing 
widths
with increasing force. This trend, if continued, would lead to a vanishing cap
 as $f \rightarrow \infty$. However, we see a different picture for $N=2$ filaments (Fig. \ref{cap-dist}b) -- the distribution $p(k)$ saturates with increasing force, implying a non-vanishing cap for multiple filaments.  
 
 This phenomenon can be understood by noting the following:
 for a multi-filament system ($N>1$), only the leading filament ``feels'' the force; the trailing filaments have force-independent rates. Therefore the trailing filaments have much higher polymerisation rates ($u_{\rm trail}=k_0 c$) compared to the leading one ($u_{\rm lead}=u_{\rm trail}\exp(-f d/k_B T)$), and hence they acquire ATP/GTP subunits at the tip. In other words, the trailing filaments ``catch up'' with the leading filaments by polymerising ATP/GTP subunits.
Thus, in a multifilament system there exists a finite cap, always, even at large forces, unlike the single filament.
%

 In summary, we have discovered a  collective phenomenon that  the  cap-sizes stabilize  with increasing force for multiple filaments, unlike a single filament, which in turn would impart enhanced  stability to multiple filaments during their catastrophes (as discussed in the last section). However, direct experimental observation of cap may be technically difficult. Hence one experimental way to observe  the above phenomenon may be the measurement of  collapse time $T_{\rm coll}$ (as discussed in the previous section). Alternatively, one may investigate  experimentally the macroscopic length fluctuations  of multi-filament system, which is quantified in the diffusion constant~\cite{fujiwara:02}. Do the length  fluctuations bear any quantitative signature of the collective effect of cap-size stabilization? We shall investigate this question in the next section.
 

\subsection*{Collective behaviour in diffusion coefficient for length fluctuations of $N$  filaments}
%
%

In this section we investigate fluctuations of the overall system-length (wall position) of an N-filament system under force, and examine plausible collective effects.
%
%
The length fluctuations can be characterised by the diffusion constant for the wall position:
%
\be
D= [\langle (\Delta x)^2 \rangle - \langle \Delta x \rangle^2 ]/2|t_2 - t_1|.
\ee
Here  $\Delta x =x(t_2)-x(t_1)$ is the difference between two  distinct instantaneous wall positions, measured at times $t_2$ and $t_1$ respectively. We calculate $D$ at the steady state ($t_1 , t_2 \rightarrow \infty$) where it is  independent of time and for the full range of forces below and above $f_s^{(N)}$.
%
%

In the literature, different groups have examined the diffusion constant for a single actin filament ($N=1$)
as a function of ATP-bound monomer concentration ($c$) at zero force~\cite{ vavylonis:05,kolomeisky:06}. It was found that $D$ has a pronounced peak near critical concentration ($c_{crit}$). This non-monotonic behaviour of $D$ was attributed to transitions between capped state and uncapped states, as a result of ATP hydrolysis. Without hydrolysis, the filament has no such transition between two states and hence $D$ is monotonic. However, the behaviour of $D$ for a multifilament system, under force, is currently unknown.
%
%
%

\begin{figure}[ht]
\centering
\includegraphics [scale=0.45]{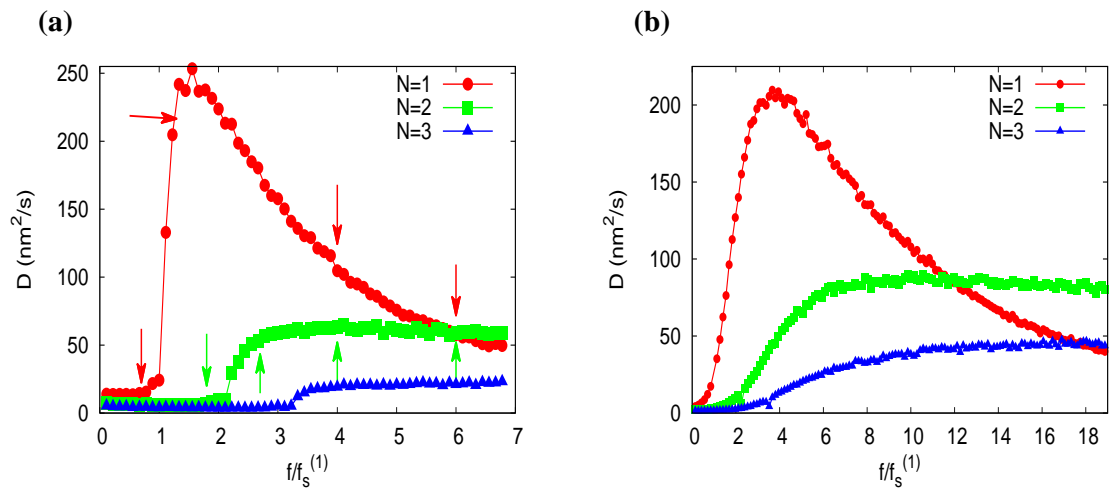}
\caption{The diffusion constant $D$  of the wall position as a function of scaled force $f/f_s^{(1)}$ for (a) actin filaments and (b) microtubules, with filament number $N=1$ (red), $N=2$ (green) and $N=3$ (blue). Concentrations are $c=0.2\mu M$ for actin and $c=10\mu M$ for microtubule (for other parameters see Table \ref{table1}). In (a), the arrows correspond to the force values at which we shall investigate the cap dynamics of the filaments  in the next section (see Fig. \ref{cap-state}). }
\label{fig4}
\end{figure}

We now present our results for diffusion coefficient $D$ in Fig. \ref{fig4}, as a function of scaled force $f/f_s^{(1)}$, both for  actin filaments (Fig. \ref{fig4}a) and microtubules (Fig. \ref{fig4}b). For one filament (red curves in Figs. \ref{fig4}a and \ref{fig4}b), we see that  $D$ rises up near the stall force $f_s^{(1)}$ and goes to zero as $f\rightarrow \infty$. Like refs. \cite{ vavylonis:05, kolomeisky:06}, we note that the non-monotonic behavior of $D$ is an effect of hydrolysis --- we have checked that this is absent for hydrolysis rate $r=0$. What is striking is that for $N>1$, $D$ curves have a distinct feature compared to $N=1$ (see green curves for $N=2$ and blue curves for $N=3$ in  Figs. \ref{fig4}a and \ref{fig4}b). For $N>1$, we see that  $D$ curves rise up near the corresponding stall forces  $f_s^{(N)}$, but they do not decay to zero at large forces like the $N=1$ case  --- in fact, they  {\it saturate} with force. As a result, the length fluctuations of a multifilament system is larger than that of a single filament system as $f\rightarrow \infty$.

The collective effect is reminiscent of the stabilization of caps with force for $N>1$ seen in the previous section. But, how exactly the microscopic dynamics of the caps contribute  to the macroscopic length fluctuation? This may be understood by examining
the transitions between ``capped'' and ``uncapped'' states of the filaments. In the next section we proceed to study these transitions as a function of applied force.



\subsection*{System length fluctuations are related to fluctuations in switching between capped and uncapped states}
%

In this section we 
demonstrate
how  transitions between capped and uncapped states of the filaments play a crucial role in the fluctuations of the wall position. To describe the instantaneous state of the tip of a single filament ($N=1$), we first define the following stochastic variable:
\bea
S(t)&=& 1, ~~\text{ if  the filament has a non-zero  ATP/GTP cap (``capped'' state)}\nonumber\\
    &=& 0,  ~~\text{if there is no ATP/GTP cap (``uncapped'' state).}
\eea
Above definition can be extended to multiple filaments. For $N>1$, we  define $S(t)=1$ or $0$ depending on whether the ``leading'' filament is capped or uncapped. Note that state of the leading filament is connected to the dynamics of the wall.
%

\begin{figure}[ht]
\centering
\includegraphics [scale=0.45]{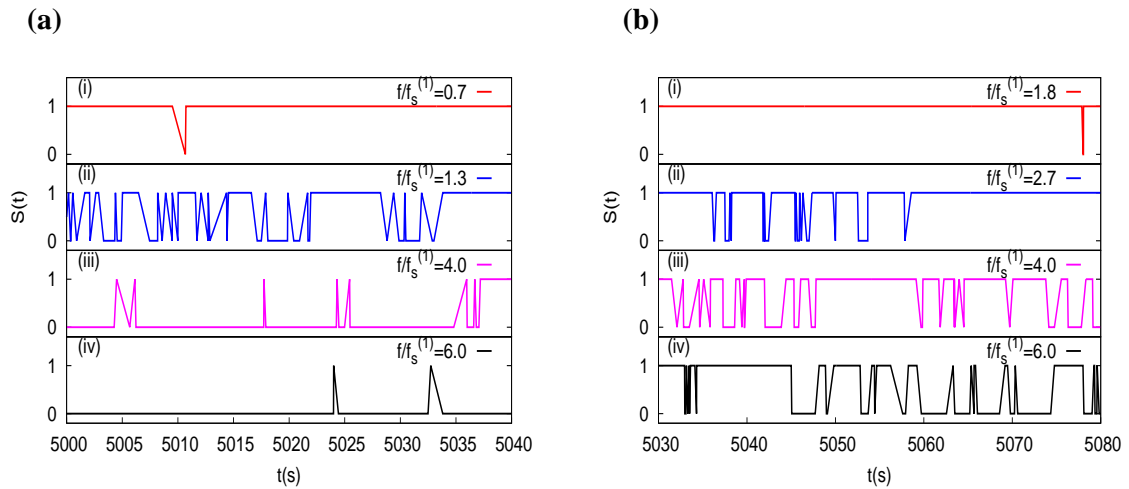}
\caption{Few time traces of  $S(t)$ of the leader for (a) $N=1$ and (b) $N=2$ actin filaments, at a concentration   $c=0.2\mu M$ and at different values of scaled forces. At these forces, the corresponding values of wall-diffusion coefficient $D$ are shown by arrows in Fig.  \ref{fig4}a (red arrows for $N=1$ and green arrows for $N=2$).}
\label{cap-state}
\end{figure}

In Fig. \ref{cap-state}a we show the time traces of $S(t)$ for a single actin filament at different force values -- at these forces, the corresponding values of wall-diffusion constant $D$ are shown by red arrows in Fig.  \ref{fig4}a. We see that, at $f \ll f_s^{(1)}$ the filament is mostly in the capped state --- $S(t)=1$ (mostly) in top panel (i) of Fig. \ref{cap-state}a.  When $f$ is just above $f_s^{(1)}$, we see in panel (ii) of Fig. \ref{cap-state}a, that there is a sudden increase in the number of switching events between capped and uncapped states. If $f$ is increased further, the number of switching events decreases --- see subsequent panels (iii) and (iv). So, the number of switching events first increases, and then decreases with force. Note that this behavior mimics the non-monotonic behavior of the wall-diffusion constant $D$, for $N=1$ (see Fig. \ref{fig4}a). Moreover, the bottom panel (iv) of Fig. \ref{cap-state}a, where $S(t)$ is mostly $0$, signifies that the filament is capless (also see Fig. \ref{ave-cap}).


We now show the time traces of $S(t)$ for two actin filaments in  Fig. \ref{cap-state}b, at different  force values; see corresponding  $D$ values in Fig.  \ref{fig4}b, marked by green arrows. Here we see a very distinct feature compared to the one-filament case  --- although the number of switching events increases first (see panels (i) and (ii) of Fig.  \ref{fig4}b), it does not decrease with force, unlike the single filament case.
%
%
%
%
In fact, the switching is present even at large forces -- compare the histories in the last panels (iv) of Figs. \ref{cap-state}a and \ref{cap-state}b. 
Furthermore, in   panels (iii) and (iv) of Fig. \ref{cap-state}b the number of switching events are nearly the same, suggesting saturation with force.  
This saturation behavior for $N>1$, may be correlated with the saturation of the wall-diffusion constant $D$ at large forces. To make this  apparent correlations between $D$ and the switching number fluctuations concrete, we now proceed to quantify the fluctuations in the number of  switching events.

\begin{figure}[ht]
\centering
\includegraphics [angle=0,scale=0.7]{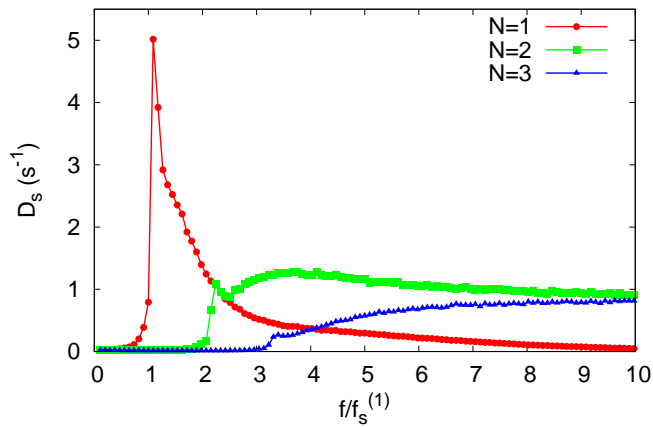}
\caption{The diffusion constant $D_s$ derived from the fluctuations of the switching events  between the capped and uncapped states, is plotted against scaled force $f/f_s^{(1)}$. The data is for actin parameters (see Table 1) at concentration  $c=0.2\mu M$, for filament-numbers $N=1$ (red),  $N=2$ (green), and $N=3$ (blue). }
\label {Ds}
\end{figure}

From the time traces of $S(t)$, we first computed the number of switching events ($n_s$) between the capped and uncapped states within a time window $\tau=|t_2 - t_1|$. We   then calculated the  variance of  $n_s$ and found that the variance grows linearly with the size of time-window i.e. ${\rm Var}[n_s] \propto \tau$. This enables us to construct a diffusion constant for the  switching events as below:
\be
D_s= \frac{1}{2}\frac{d}{d\tau}\rm Var[n_s].
\ee
 We  compute $D_s$ at large times, where it becomes independent of time.

In Fig. \ref{Ds} we plot $D_s$ versus $f/f_s^{(1)}$ for actin parameters (see Table 1). Quite strikingly, we see that behavior of $D_s$ is very similar to the behavior of wall-diffusion constant $D$ (see Fig. \ref{fig4}a). Just like the wall-diffusion constant,  at large forces,  $D_s$ goes to zero for $N=1$, and it saturates for $N>1$. This clearly demonstrates that the wall-position fluctuations (quantified by $D$) are closely tied  to the fluctuations of the switching events (quantified by $D_s$) between the capped and uncapped states. 

\section*{Discussion and Conclusion}

The current understanding of dynamical properties and fluctuations of cytoskeletal filaments, with hydrolysis, is mostly based on studies of single filaments \cite{hill:81, kolomeisky:06, Leibler-cap:96,vavylonis:05,Ranjith2009,Ranjith2010,Ranjith2012}. Recent experiments by Laan et al \cite{Laan-pnas:08} and subsequent theory papers have started exploring various aspects of multiple filament systems under force \cite{Kierfeld:2013,das:arxiv}. However, there is no clear understanding of mechanisms leading to catastrophe, length fluctuations and cap dynamics of a mutifilament system. In this paper, using a detailed stochastic model of multiple filaments under force,  taking into account polymerisation, ATP/GTP hydrolysis and depolymerisation of T- and D-bound subunits, we systematically investigated and clarified a number of aspects related to the dynamics and fluctuations of the system. Specifically, we showed that the fluctuations during collective catastrophes,  the fluctuations of the ATP/GTP cap sizes, and the system length fluctuations, all are closely tied to each other. The unified picture emerging from these studies show that the collective behaviour of multiple filaments are quantitatively distinct from that of a single filament under similar conditions. For example, multifilament systems are more stable during catastrophe, when compared to a single filament system. Thus, our study suggests that it would be inaccurate to generalise the intuitions built on existing studies of single filaments to the more biologically relevant scenario of multiple filaments.

We quantified the fluctuations during  catastrophes  by  the mean collapse time ($T_{\rm coll}$). We  found that $T_{\rm coll}$ is systematically lower for a single filament compared to multiple filaments. This implies that the multi-filament system has an enhanced resistance  against externally applied force. This will also clearly reflect in the experimentally measurable length versus time data, where, according to our prediction, the collective collapse will have a lower average negative slope, unlike the sharp collapse which is the hallmark of a single filament catastrophe \cite{libchaber_PRE:94,rob-phillips:book}. Recent experiments on multiple microtubules under force clearly show this trend of slower collapse in their length versus time data (see Ref. \cite{Laan-pnas:08}, Fig. 2A). We would like to note that this interesting feature, an understanding of which naturally emerges from our model, seems to be absent in time traces of wall positions obtained using other theoretical models in the literature (models in \cite{Laan-pnas:08,Kierfeld:2013}).

Exploring the ATP/GTP cap structure and statistics of individual filaments in the multifilament system, we found those to be highly stable at large forces.
This enhanced stability of the caps (for $N>1$) imparts  stability to a multi-filament system, which is responsible for their  slow collapse. Moreover, the stability of the caps is also reflected in the macroscopic length fluctuations of $N$ filaments, which we quantified by a diffusion constant ($D$). We find that, at large forces, the value of $D$ (for $N>1$) saturates -- this experimentally observable effect owes its origin to the number fluctuations of the switching events between the capped and uncapped states (quantified by $D_s$). The similarity of the curves of $D$ and $D_s$ (versus force) demonstrates this. (see Figs. \ref{fig4}a and \ref{Ds}).
In single microtubule dynamics, presence of GTP-bound subunits in the bulk is associated with rescue \cite{perez:2008}. In multifilament systems one would expect enhanced rescues, at smaller forces closer to the stall force, as  the lagging filaments can easily acquire GTP-bound subunits.

Although, due to technical difficulties, the cap may not be directly observable experimentally, 
other quantities like the collapse time $T_{\rm coll}$ and the diffusion constant $D$ can be measured in a laboratory. Note that
 our definition of $T_{\rm coll}$ and $D$ rely on just the time traces of the system length, which can be obtained easily in well designed experiments. It is worth mentioning that $T_{\rm coll}$ may be used to determine the stall force of a multifilament system and its deviation from the additive law (i.e. $f_s^{(N)}>N f_s^{(1)}$), as predicted in our earlier work \cite{das:arxiv} can be verified.
 
We would like to conclude by pointing out that dynamics of cytoskeletal filaments under any situation providing a scope for cooperativity (e.g, a boundary wall held by a force, as in our case) may produce surprises for multi filaments, and understanding of such situations should start with case studies of at least two filaments. Any conclusion based on single filament study, in such cases, would be misleading.

\subsection*{Acknowledgments}
We acknowledge CSIR India (Dipjyoti Das, JRF award no.~09/087(0572)/2009-EMR-I) and IYBA, Department of Biotechnology India 
(RP, No: BT/01/1YBAl2009) for financial support.

\appendix
\def\figurename{Fig. A{\hskip -2.5pt}}
\def\tablename{Table. A{\hskip -2.5pt}}
\setcounter{figure}{0}
\setcounter{table}{0}

\section*{Appendix A: Different phases of growth and shrinkage}

\begin{figure}[]
\centering
\includegraphics [scale=0.45]{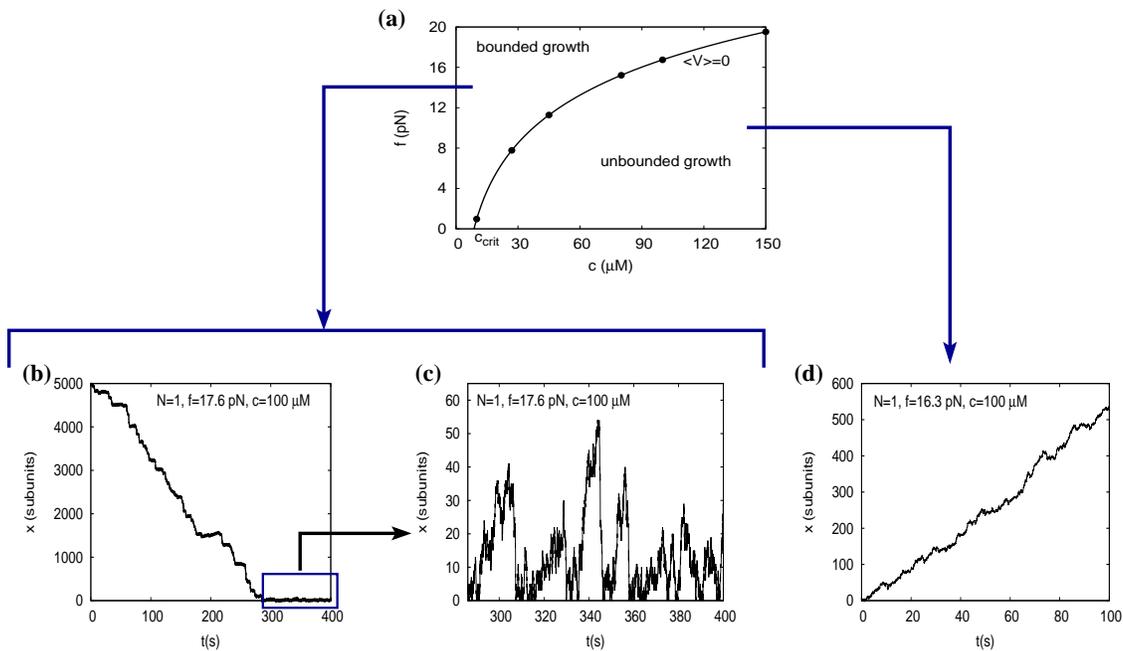}
\caption{(a) Phase diagram of $N=1$ microtubule in the force (f)-concentration (c) plane. The curve of mean wall-velocity $\langle V \rangle =0$ demarcates between two phases, namely the bounded and unbounded growth phases. (b) and (c): Typical time traces of the wall position in the bounded phase. The trajectory of (b) shows that the system length (wall position) $x(t)$ first shrinks rapidly with a negative velocity, but ultimately it fluctuates around a constant mean value --- the later part is zoomed in (c), which shows catastrophes of the filament. (d) A typical trajectory of the system length in the unbounded growth phase, where $x(t)$ grows in time with a positive velocity. Parameters are specified in Table \ref{table1} and inside the figure panels. }
\label{phases}
\end{figure}

Depending on the values of the applied force $f$ and concentration $c$, there are two dynamical phases of a $N$-filament system --- (i) the bounded growth phase, and (ii) the unbounded growth phase \cite{Leibler:93,Ranjith2012}. The phase diagram for $N=1$ microtubule is shown in the $f-c$ plane, in Fig. A\ref{phases}a. The $\langle V \rangle =0$ curve marks the phase-boundary (where, $\langle V \rangle$ is the mean wall velocity). It should be noted that, in the absence of force ($f=0$), there exists a critical ATP/GTP concentration $c=c_{crit}$, at which $\langle V \rangle=0$ i.e. the system is stalled (see Fig. A\ref{phases}a). In the presence of force ($f>0$), and for a concentration $c>c_{crit}$, the system can only be stalled when we apply the  ``stall force'', $f=f_s^{(N)}$ at which the average wall velocity $\langle V \rangle = 0$ \cite{lacoste11,das:arxiv}. 

For the parameter regime   $c>c_{crit}$ with $f>f_s^{(N)}$, or for $c<c_{crit}$ with $f\geq 0$, the filament shrinks on an average with a negative velocity --- see the trajectory of the wall position in Fig. A\ref{phases}b. During shrinkage, when the filament length becomes very short,  the filament eventually encounters the immovable left wall, and then  the length  fluctuates around a constant mean value --- see Fig. A\ref{phases}c. This is called the bounded phase. On the other hand, for the parameter regime  $c>c_{crit}$ and $f<f_s^{(N)}$, the filament indefinitely grows on an average, with a positive velocity --- this is the unbounded growth phase --- see a typical trajectory within this phase in Fig. A\ref{phases}d. 

\section*{Appendix B: Comparison of stall forces obtained from the collapse time measurements, and from the force-velocity relations}

In this appendix, we discuss two possible measurement procedures for the stall force. 
In theories, the stall force is usually measured from force-velocity relations by finding the force $f=f_s^{(N)}$, at which average wall velocity $\langle V(f) \rangle = 0$ \cite{lacoste11,Ranjith2009,van-doorn:00}. But a fact is that $\langle V \rangle$ becomes very small for $f \rightarrow f_s^{(N)}$, 
with increasing $N$. As a result, for $N > 2$, the task of experimentally measuring the precise $f_s^{(N)}$ is challenging. In fact, from simulations we calculated that for $N = 3$ microtubules (or actin filaments), at a force $f=3 f_s^{(1)}$ and concentration  $c=100 \mu M$ (or $c=1 \mu M$),  the wall velocity is $\sim 0.1$ nm/s (or $\sim 0.02$ nm/s).  Monitoring such slow motion, and finding the force for which the wall truly halts maybe  difficult. 

We have already shown  in the main text that, if $f \rightarrow f_s^{(N)}$ from above, the average collapse time $T_{\rm coll}$  tends to diverge. This behavior of $T_{\rm coll}$  may be used to precisely determine the collective stall force  $f_s^{(N)}$ of a system. This relies on approaching $f_s^{(N)}$ from above, as opposed to from below as in the case of $f-\langle V \rangle$ measurements.  Note that such a measurement of $T_{\rm coll}$ should be easy experimentally, as one is dealing with large values, while monitoring vanishingly small values of $\langle V \rangle$ is more difficult.   We present the values of $f_s^{(N)}$ obtained numerically by noticing the limits  $\langle V \rangle \rightarrow 0$, and $T_{\rm coll} \rightarrow \infty$, in Table A\ref{table2} --- they match quite well.  

\begin{table}[]
\caption{Comparison of values of stall forces obtained numerically by monitoring the limits  $\langle V \rangle \rightarrow 0$, and $T_{\rm coll} \rightarrow \infty$. ATP/GTP Concentrations are taken to be $c=1\mu M$ for actin and $c=100\mu M$ for microtubule (for other parameters see Table \ref{table1}).}
\vspace{2mm}
\label{table2}
\centering
\begin{tabular}{|c|c|c|c|c|c|c|} \hline
~  & & $f_s^{(1)}$ (pN) &  $f_s^{(2)}$ (pN) & $f_s^{(3)}$ (pN)& $f_s^{(4)}$ (pN) \\ \hline
Actin & $\langle V \rangle$ measurement & $3.134$ & $6.389$ & $ 9.619$  & $12.834$ \\ \cline{2-6}
& $T_{\rm coll}$ measurement & $3.134$ & $6.390$ & $9.616$  & $12.832$ \\ \hline \hline
MT & $\langle V \rangle$ measurement & $16.748$ & $35.010$ & $52.793$  & $70.384$ \\ \cline{2-6}
 & $T_{\rm coll}$ measurement & $16.741$ & $35.017$ & $52.814$  & $70.473$ \\ \hline
\end{tabular}
\end{table}
\newpage
\bibliography{fluctuation-paper-BJ-to-submit}
\end{document}